\newcommand{\be}{\begin{eqnarray}}
\newcommand{\ee}{\end{eqnarray}}
\newcommand{\nn}{\nonumber}

\documentclass[12pt]{article}

\usepackage{amsmath,amssymb}
\usepackage{graphicx}
\usepackage{cite}
\usepackage{bm}
\begin{document}
\renewcommand{\thefootnote}{\fnsymbol{footnote}}

\vskip 15mm

\begin{center}

{\Large On  dynamics of $5D$ superconformal theories}

\vskip 4ex

A.V. \textsc{Smilga}

\vspace{.5cm}

SUBATECH, Universit\'e de
Nantes,  4 rue Alfred Kastler, BP 20722, Nantes  44307, France
\footnote{On leave of absence from ITEP, Moscow, Russia.}
 
$\texttt{smilga@subatech.in2p3.fr}$
\end{center}

\vskip 5ex

\begin{abstract}
\noindent   $5D$ superconformal theories involve  vacuum valleys characterized in the simplest case by the vacuum expectation value
of the real scalar field $\sigma$.
If $\langle \sigma \rangle \neq 0$, conformal invariance is spontaneously broken and 
the theory is not renormalizable. In the conformally invariant sector   $\langle \sigma \rangle = 0$, the theory is 
intrinsically nonperturbative. We study classical and quantum dynamics of this theory in the limit when field dependence of the 
spatial coordinates is disregarded. 
The classical trajectories ``fall'' on the singularity at $\sigma = 0$. The  quantum spectrum involves ghost states with unbounded from below
negative energies, but such states fail to form complete 16-plets as is dictated by the presence of four complex supercharges and should
be rejected by that reason. Physical excited states come in supermultiplets and have all positive energies. 
 We conjecture that the spectrum of the complete field theory hamiltonian is nontrivial and has a similar nontrivial ghost-free 
structure and also  
 speculate that the ghosts in higher-derivative supersymmetric  field theories are exterminated by a similar mechanism.
\end{abstract}

\renewcommand{\thefootnote}{\arabic{footnote}}
\setcounter{footnote}0
\setcounter{page}{1}

\section{Introduction}
 
Field theories in more than four dimensions attracted recently a considerable attention. Usually they are discussed
in the string theory perspective, but to our mind higher dimensional theories are interesting {\it per se}. 
In particular, certain arguments may be given  \cite{TOE} that a variant of higher-derivative 
superconformal theory in higher dimensions 
(possibly, a theory enjoyng the maximal ${\cal N} =2$ superconformal symmetry in
 six dimensions) may  be the fundamental Theory of Everything.  

For a field theory to make sense, the continuum limit of the path integral when the lattice spacing is sent to zero or, 
what is equivalent, the ultraviolet cutoff $\Lambda$ is sent to infinity should be well defined.
Usually,  this property is associated with renormalizability. Indeed, a renormalizable theory is a theory with
a { finite}
 number of bare couplings defined at ultraviolet scale $\Lambda$, which can be sent to infinity in such a way that the effective
 lagrangian describing the physics at energies $E \ll \Lambda$ also involves a finite number of couplings  
 that do not depend on $\Lambda$.
 Renormalizable theories are abundant in 4 dimensions. They are divided in two types: {\it (i)} 
asymptotically free and conformal 
 theories with nontrivial continuum limits and {\it (ii)} the theories like ordinary QED, where  the only meaningful 
continuum limit describes a free  theory.

It is difficult, however, to construct a sensible renormalizable theory for $D > 4$. For example, 
a usual YM theory with the action
$\sim {\rm Tr} \{F_{\mu\nu} F_{\mu\nu} \}$ involves in higher dimensions a dimensionful coupling and  is not renormalizable.
If one wishes still to work with higher-dimensional theories, one should choose between the following nonstandard options:

First, one can  remark that  renormalizability is a statement about the benign perturbative structure of the theory. 
Though it is {\it sufficient} for the continuum limit to exist, we do not know whether it is also {\it necessary}.
Even in perturbative framework, one can imagine a theory involving an infinite number of counterterms which absorb all power
ultraviolet divergences so that the physical quantities do not depend on $\Lambda$ \cite{nonren}.  One such theory has actually
been studied for a long time, I mean the chiral perturbation theory \cite{ch}. A distinguishable property of the latter is that loops
involve there not only power UV divergences cancelled out by counterterms, but also so called chiral logarithms which lead
to nontrivial contributions $\propto \ln (\mu_{\rm hadr}/m_\pi)$ in all physicall quantities. 
 For sure, chiral theory is an effective theory describing low-energy dynamics of QCD and  not a fundamental theory. But we cannot exclude
at present that a fundamental theory with nonrenormalized lagrangian that leads to nontrivial cutoff-independent dynamics at all energy
scales can be defined.

Another possibility is to keep renormalizability by adding  higher derivatives in the lagrangian. In \cite{ISZ}, we constructed 
a renormalizable $6D$ higher-derivative theory. It is a supersymmetric gauge theory with the terms 
$\propto (D_\mu F_{\mu\nu} )^2$ etc. in the lagrangian so that the coupling constant is dimensionless and only logarithmic 
ultraviolet divergences (handled by the renormalization procedure) are present. At the classical level, this theory
enjoys conformal symmetry, which is broken, however, by quantum anomaly. The effective charge runs there increasing with the
 energy  similar to what happens in ordinary $4D$ QED.

Higher derivative theories have been seldom considered seriously by theorists and the reason is the problem of ghosts. It is very difficult
to get rid of the ghosts (physically, they mean instability of the perturbative vacuum and are associated with 
the absence of the ground state in the hamiltonian) when the lagrangian involves higher derivatives. In \cite{duhi} we 
managed, however, to construct a quantum mechanical example where higher derivatives are present and still the ground 
state is  well defined. This and some other observations encouraged us to speculate in \cite{TOE} that the TOE 
may represent a $6D$ superconformal higher-derivative theory. The example considered in \cite{duhi} was not supersymmetric. However,
(and this is one of the observations of the present paper), it is much easier to cope with ghosts in a supersymmetric theory 
than in a non-supersymmetric one. Indeed, supersymmetry usually implies positivity of all energies (only the energy of vacuum can be zero).
And this means that the spectrum is bounded from below and ghosts are absent. We return to the discussion of this question in the end of the paper.

Besides two possibilities mentioned above ({\it (i)}  nonrenormalizable theory where a nontrivial continuum limit exists 
and {\it (ii)}   renormalizable theory with higher derivatives), there is also a third possibility to obtain a viable extra dimensional theory.
A theory can be {\it intrinsically nonperturbative} so that even a question whether it is renormalizable or not cannot be posed. An example
of such theory is superconformal theory in five dimensions.
 This theory was first discussed  qualitatively in \cite{Seiberg}, and its action was written    in the language of harmonic superfields
in \cite{Zup} and in the component form in \cite{5Dcomp}. 
The simplest such theory involves an Abelian gauge field $A_\mu$, a scalar field $\sigma$, auxiliary fields 
 and fermionic superpartners. The lagrangian  is
\footnote{Our conventions are here the following. We use the metric $\eta_{\mu\nu} = {\rm diag}(1,-1,-1,-1,-1)$. The $\gamma$-matrices 
$\tilde \gamma_\mu$ (we put tildas to distiguish these $\gamma$ matrices from the untilded Euclidean $\gamma$ matrices to be introduced later) satisfy 
$\tilde \gamma_\mu \tilde\gamma_\nu + \tilde\gamma_\nu \tilde\gamma_\mu = 2\eta_{\mu\nu}$, $\tilde\gamma^\dagger_0  = 
 \tilde\gamma_0, \tilde\gamma^\dagger_{1,2,3,4} = -\tilde\gamma_{1,2,3,4}$. 
The usual notation  $\tilde\sigma_{\mu\nu} = \frac 12 (\tilde\gamma_\mu\tilde \gamma_\nu - \tilde\gamma_\nu\tilde\gamma_\mu)$ is used.
 The charge conjugation or symplectic matrix $\tilde C$ 
 (the algebra $Spin(4,1)$ is equivalent up to irrelevant complexities to $Sp(4)$)  satisfies the properties
$$ \tilde C^T = -\tilde C, \ \ \ \ \ \ \ \ \tilde C^2 = -1\!\!\, \ \ \ \ \ \ \ \tilde C\tilde \gamma_\mu^T = \tilde\gamma_\mu \tilde C$$
  so that  the form
$\tilde C^{ab} \psi_a \chi_b$ is invariant with respect to Lorentz transformations for arbitrary spinors $\psi_a, \chi_b$. One of the possible explicit
choices for $\tilde\gamma_\mu$ and $\tilde C$ is
 $$ \tilde\gamma_0 = 1\!\!1 \otimes \sigma_1,\ \ \ \ \ \ \tilde\gamma_{1,2,3} = i\sigma_{1,2,3} \otimes \sigma_2, \ \ \ \ \ \tilde\gamma_4 = -i 1\!\!1 \otimes
\sigma_3,\ \ \ \ \ \ \tilde C = i\sigma_2 \otimes 1\!\!1\ . $$
The convention for the auxiliary fields 
$D_{jk}$ follows Ref.\cite{ISZ} and is such that $D_{12} = D_{21}$ are real and $D_{11} = -D_{22}^*$. Then $D^{jk} D_{jk} < 0$. }
 \be
\label{L5D} 
  g^2{\cal L}\ =\ -\frac \sigma 4 F_{\mu\nu} F_{\mu\nu} + \frac \sigma 2 (\partial_\alpha \sigma)^2 +
\frac {i\sigma} 2 \bar \psi^j /\!\!\!\partial \psi_j - \sigma D^{jk} D_{jk} \nn \\
\frac i8 \bar \psi^j \tilde\sigma_{\mu\nu} F_{\mu\nu} \psi_j + \frac 1{24} \epsilon_{\mu\nu\lambda\rho\sigma} A_\mu F_{\nu\lambda} F_{\rho\sigma}
+ \frac 12 \bar \psi^j \psi^k D_{jk}\, ,
 \ee 
where 
 $\psi_j$ is the pseudo-Majorana fermion satisfying the constraint
 \be
\label{constr}
\bar \psi^{ia} \equiv (\psi_{ia})^* \tilde\gamma^0 = \epsilon^{ij} \tilde C^{ab} \psi_{jb}\, .
 \ee

\section{Field theory: perturbative and nonperturbative  features.}
 
The lagrangian (\ref{L5D}) does not involve higher derivatives. Still, it is scale (actually, conformally) 
invariant and the coupling constant $g^2$ is dimensionless. 
The latter would suggest renormalizability of the theory. However, the lagrangian (\ref{L5D}) does not involve a quadratic part and 
perturbative calculations
based on smallness of the interaction term with respect to the free part are impossible ! In this respect,  the situation  is even worse than in a
nonrenormalizable theory. In the latter, perturbative series diverges, but one can at least go into the interaction representation, 
define asymptotic scattering states,  and if not to find 
what their scattering matrix  is, but at least  ask this question. Here we seem not to be able to  do it .

An option that one has, however, is to  capitalize on the 
presence  of the vacuum moduli space in the lagrangian (\ref{L5D}). Indeed, any vacuum expectation value $\langle \sigma \rangle = m$ 
is equally admissible and does not cost energy. If $m \neq 0$, we can pose $\sigma = m + \sigma'$ treating $\sigma'$ perturbatively.
 In this case, conformal symmetry 
of the original lagrangian is broken spontaneously and the scale parameter $m$ is introduced. This allows one to decompose the lagrangian into
the free and interaction parts and  construct the $S$-matrix in a conventional way.
Inverting the quadratic part of the lagrangian, we obtain the propagators 
 \be
\label{prop}
 \langle \sigma \sigma \rangle \ =\ \frac i{mp^2}\, , \nn \\
\langle A_\mu A_\nu \rangle \ =\ - \frac {i\eta_{\mu\nu}}{mp^2} \, ,   \nn \\
 \langle \psi_j \bar \psi^k \rangle \ =\ \delta_j^k \frac {i /\!\!\!p}{mp^2}\, .
 \ee
The mass parameter in the denominator makes the theory nonrenormalizable. In particular, the vacuum expectation value $\langle \sigma \rangle$
does not want to keep its bare value $m$, but involves power divergent corrections. 
 
The easiest way to find the one-loop correction to $\langle \sigma \rangle$ is 
 to consider the
term $\propto D^{jk} D_{jk}$ in the effective lagrangian. Only one graph depicted in Fig. 1 contributes. The calculation gives
 \be
\label{DelL}
\Delta {\cal L}_{\rm eff} =  -\frac { D^{jk} D_{jk}}{m^2} \int \frac {d^5p_E}{(2\pi)^5 p_E^2} \equiv 
 -\frac { D^{jk} D_{jk}}{m^2} \Lambda^3\ .
 \ee
\vspace{.7cm}

\begin{figure}[h]
   \begin{center}
 \includegraphics[width=3.0in]{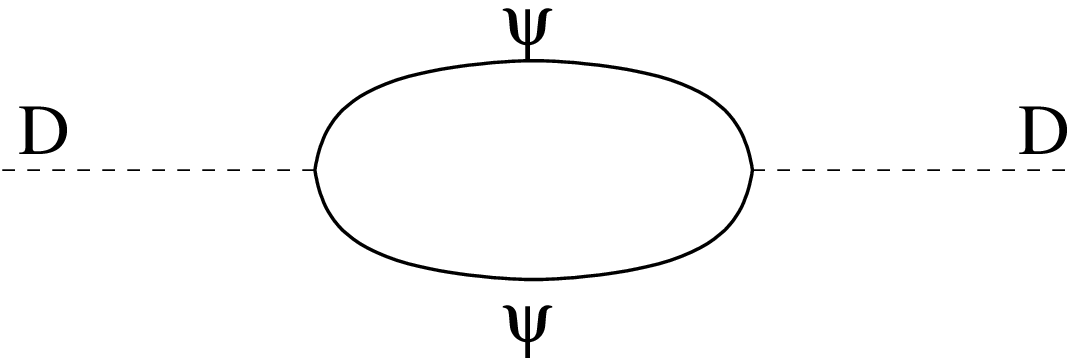}
        \vspace{-2mm}
    \end{center}
\caption{}
\label{D5D}
\end{figure}

In other words, the vacuum expectation value acquires a  shift involving a  cubic ultraviolet divergence
 \be
\label{1loop}
 \langle \sigma \rangle_{\rm 1\  loop} \ =\ m + \frac {\Lambda^3}{m^2}
 \ee
The two-loop correction is expected to be of order $\Lambda^5/m^4$, which is much larger than $\Lambda^3/m^2 \gg m$, etc. 
 In addition, higher-dimensional operators in the effective lagrangian are generated, which should in turn be added to the tree lagrangian.
  The situation is the same as in any other nonrenormalizable field theory. Maybe one can handle it, as  discussed above, by adding an infinite number
of fine-tuned counterterms, but we are not able to say anything more in this respect.

But what  if  $\langle \sigma \rangle = 0$ ?  In this case, one cannot treat the theory perturbatively whatsoever. 
This does not mean, however, that the theory does not have a nonperturbative cutoff-independent meaning. It is quite possible that its
 hamiltonian has well defined ground state and excitations above it.
Unfortunately, we do not have the means to tackle a nonlinear quantum field theory problem analytically and the guess above can
neither be substantiated not disproved. Let us do few things which we can. To begin with, one can wonder what is the classical dynamics
of the lagrangian (\ref{L5D}). Let us limit ourselves by the scalar sector. The equations of motion read $\Box \sigma^{3/2} = 0$ with the solution
  \be
 \label{sigwave}
 \sigma(x_\mu) \ =\ \left[ \cos(kx + \delta) \right]^{2/3}
 \ee
with $k^2 = 0$. This is singular at $kx + \delta = \pi/2 + \pi m$. 

One can compare
this theory to ordinary hydrodynamics. If viscosity is zero, the latter can be cast into lagrangian (or hamiltonian) form, but the quadratic terms
in such a hamiltonian (formulated in terms of so called Klebsch variables) are absent. A typical solution to the Euler equations of motion is also singular.
Smooth laminar solutions exist, but they are unstable with respect to creation of vortices. The scale of these vortices 
 becomes rapidly very small leading to large values of the velocity gradients etc. In reality, 
the minimal size of the vortices is controlled by the viscosity playing the role of ultraviolet cutoff. But in the mathematical limit
of zero viscosity, classical solutions run into singularity. 
The latter also happens in simple mechanical systems with strong attractive potential.

 In our case, the situation is somewhat worse because in hydrodynamics or in the mechanical system with the potential 
$V(r) = -\gamma/r^2$, only {\it some} of the classical trajectories are singular while regular trajectories (laminar motions in hydrodynamics
and the trajectories with large enough angular momentum for the problem with 
$V(r) = -\gamma/r^2$ ) also exist. In the theory under consideration,  {\it all}  classical solutions seem to be 
singular. In other words, the classical theory with the lagrangian (\ref{L5D}) seems to be meaningless.

This is not a final diagnosis yet as classically meaningless (singular) theories {\it can} acquire meaning when quantized. A classical
example is the QM theory with $V(r) = - \gamma/r^2$. If $\gamma < 1/(4m)$, the spectrum of the quantum hamiltonian is well defined in spite of the fact 
that its classical counterpart involves singular trajectories. Another example was considered in \cite{duhi} where 
we showed that a higher-derivative mechanical system involving
singular classical trajectories  has a well-defined quantum spectrum. 

Is it also the case for the theory (\ref{L5D}) ?

 \section{Quantum mechanics.}

In this section, we study  nonperturbative quantum dynamics of the lagrangian (\ref{L5D}) dimensionally reduced in $(0+1)$ dimensions (the only
limit when we can do it). The reduced lagrangian involves five dynamical bosonic degrees of freedom (four spatial components of $A_\mu$ and
$\sigma$). After suppressing spatial dependence and  excluding auxiliary fields $D_{jk}$ the lagrangian acquires the form
 \be
\label{Lpredel}
L\ =\ \frac \sigma 2 \left( \dot A_K^2 + \dot \sigma^2  + i \psi^{\dagger j} \dot \psi_j \right) \nn \\
- \frac i4 \psi^{\dagger j} \tilde\gamma_K \psi_j \dot A_K + \frac 1{16\sigma} \left( \psi^{\dagger j} \tilde\gamma_0 \psi^k \right)
 \left( \psi^\dagger_j \tilde\gamma_0 \psi_k \right)
 \ee
with $K = 1,2,3,4$. 
This QM lagrangian involves  four complex conserved supercharges in virtue of its 5D origin. It belongs to the class of ${\cal N} =4$ SQM 
lagrangians suggested in \cite{DE} and studied in \cite{DEja,eshche}. A generic such lagrangian in component notation is \cite{DEja}
 \footnote{We corrected the sign error in the second term in Eq.(2.14) of Ref.\cite{DEja}.} 
\be
 \label{DEcomp}
L = \ h \left[ \frac 12 \dot A_J^2 + \frac i2 \left( \eta^\dagger  \dot \eta -  \dot { \eta^\dagger} \eta \right) \right]
 - \frac i2 (\partial_K h) \dot A_J \eta^\dagger
 \sigma_{KJ} \eta \nn \\
+ \frac 1{24} \left[ 2 \partial_J \partial_K h - \frac 3h (\partial_J h) (\partial_K h) \right] \left[
(\eta^\dagger \gamma_J \eta )(\eta^\dagger   \gamma_K \eta) - (\eta  C  \gamma_J \eta) (\eta^\dagger 
\gamma_K C \eta^\dagger) \right]  \ ,
 \ee
where $J = 1,\ldots 5$ with Euclidean hermitian $\gamma_J$, and $h(A_J)$ is a 5-dimensional 
harmonic function, $\partial_J^2 h = 0$. This lagrangian is invariant up to a total derivative with respect 
to the following SUSY transformations,
 \be
\label{susytran}
 \delta A_J &=& \eta^\dagger \gamma_J \epsilon + \epsilon^\dagger  \gamma_J \eta\ , \\ \nonumber 
 \delta \eta_\alpha &=& -i \dot A_J ( \gamma_J \epsilon)_\alpha + 
\frac {\partial_J h}{2h} \eta^\dagger  \gamma_J \eta \, \epsilon_\alpha +
 \frac {\partial_J h}{2h}   (\eta  C \gamma_J \eta) (\epsilon^\dagger  C)_\alpha\ ,  \\ \nonumber 
\delta \eta^{\dagger \beta} &=& i \dot A_J (\epsilon^\dagger \gamma_J )^\beta + 
\frac {\partial_J h}{2h} \eta^\dagger \gamma_J \eta \, \epsilon^{\dagger \beta} + 
   \frac {\partial_J h}{2h}  (\eta^\dagger  \gamma_J  C \eta^\dagger) ( C \epsilon)^\beta \ .
  \ee
It is not difficult 
to see that   (\ref{DEcomp}) goes over into (\ref{Lpredel}) if taking  $h(A_J) = A_5$ (obviously, it is harmonic),
 if identifying 
 \be
 \label{sootv}
 A_5 \equiv \sigma \, ,\ \ \ 
\eta  \equiv \psi_1 \, ,\ \ \  \gamma_5 \equiv \tilde \gamma_0\, , \ \ \   
 \gamma_{1,2,3,4} \equiv  \tilde \gamma_0  \tilde \gamma_{1,2,3,4} ,\ \ \  \ C \equiv   \tilde \gamma_0  \tilde C
 \ee 
and using the property (\ref{constr}) with the corollary $C \eta^\dagger = -\eta^\dagger C  \equiv \psi_2$ . 
The metric $ds^2 = \sigma (d\sigma^2 + dA_K^2)$ describes an orbifold of  conic variety,  though it 
 does not represent a flat cone but has a nontrivial curvature. At the ``tip of the cone'' (  $\sigma= 0$) 
the curvature is singular, $R \sim 1/\sigma^3$.

 The canonic hamiltonian can be derived from (\ref{Lpredel}), (\ref{DEcomp}) by a usual procedure. 
It is convenient to introduce $\xi = h^{1/2} \eta$
such that $\xi^\dagger$ is canonically conjugate to $\xi$ and the Poisson bracket  $\{\xi^\dagger, \xi\}_{\rm PB}$ equals to 1. 
In a generic case, we have
 \be
\label{HDE}
 H \ =\ \frac 1{2h} \left[P_J + \frac {i\partial_K h}{2h} \xi^\dagger  \sigma_{KJ} \xi \right]^2  \nn \\ 
-   \frac 1{24h^2} \left[ 2 \partial_J \partial_K h - \frac 3h (\partial_J h) (\partial_K h) \right] \left[
(\xi^\dagger  \gamma_J \xi )(\xi^\dagger   \gamma_K \xi) - (\xi C  \gamma_J \xi) 
(\xi^\dagger  \gamma_K C \xi^\dagger) \right]\, , 
 \ee
where $P_J$ is the canonic momentum 
 \be
\label{PJ}
 P_J \ =\ h\dot A_J - \frac i2 (\partial_K h) \eta^\dagger \sigma_{KJ} \eta\ .
 \ee
 The classical supercharges are obtained in a relatively direct though  tedious way (see Appendix for some details)
by the standard N\"other procedure. The result is
  \be
\label{QDE}
Q_\alpha &=&  f P_I (\gamma_I \xi)_\alpha   + \frac i2 Q_{IJ} (\partial_J f) ( \gamma_I \xi)_\alpha  
-  \frac i{24} \epsilon_{IJKLM} (\partial_I f)  Q_{JK} (\sigma_{LM} \xi)_\alpha\ , 
  \nn \\
Q^{\dagger\beta} &=&   f P_I (\xi^\dagger  \gamma_I)^\beta    + \frac i2 Q_{IJ} (\partial_J f)  (\xi^\dagger  \gamma_I)^\beta  
+  \frac i{24} \epsilon_{IJKLM} (\partial_I f)  Q_{JK} (\xi^\dagger \sigma_{LM})^\beta \ , 
 \ee
where we introduced $f = 1/\sqrt{h}$, $Q_{IJ} = \xi^\dagger  \sigma_{IJ} \xi$.
By construction, they satisfy the  supersymmetry algebra 
 \be
\label{clalg}
\{Q_\alpha, Q_\beta \}_{\rm P.B.} = \{Q^{\dagger \alpha}, Q^{\dagger\beta} \}_{\rm P.B.} = 0\ , \ \ \ \ \ \ \ \ \ \ 
\{Q^{\dagger\beta}, Q_\alpha\}_{\rm PB} = 2\delta_\alpha^\beta H\, .
 \ee
To determine quantum supercharges, one has to resolve ordering ambiguities such that the classical SUSY algebra (\ref{clalg}) 
is kept intact at the quantum level. 
An universal receipt to do it is to use Weyl ordering for supercharges \cite{howto}, i.e. to write the quantum operators 
$\hat Q_\alpha$ and $\hat{\bar Q}^\beta$
in such a way that their Weyl symbols  would coincide with the classical expressions (\ref{QDE}). 

The proof of this statement is simple. Note that the Weyl symbol of the anticommutator of supercharges coincides with the
Moyal bracket of their Weyl symbols. The Moyal bracket introduced in \cite{Moyal} was generalized on the case when the phase space
involves not only bosonic $(P_J, A_J)$, but also canonically conjugate  fermionic 
variables $(\xi^{\dagger \alpha}, \xi_\alpha)$  in Ref.\cite{howto}.
The definition is
 \be
\label{Moyal}
 i\{A,B\}_{\rm M.B.} \ =\ 2\sinh \left\{ \frac 12 \sum_\alpha 
\left( \frac {\partial^2}{\partial \xi_\alpha^{(2)} \partial \xi^{\dagger \alpha (1)}} -   
  \frac {\partial^2}{\partial \xi_\alpha^{(1)} \partial \xi^{\dagger \alpha (2)}} \right)  \right. \nn \\
 \left. - \frac i2 \sum_J \left(
\frac {\partial^2}{\partial A_J^{(2)} \partial P_J^{ (1)}} - \frac {\partial^2}{\partial A_J^{(1)} \partial P_J^{ (2)}} 
\right) \right\} \nn \\
\left.  A \left (P_J^{(1)}, A_J^{(1)}; \xi^{\dagger \alpha\, (1)}, \xi^{(1)}_\alpha \right)
  B \left (P_J^{(2)}, A_J^{(2)}; \xi^{\dagger \alpha\, (2)}, \xi^{(2)}_\alpha \right) \right|_{1=2}\ .
 \ee
 In our case, besides the first term of the expansion of hyperbolic sine 
 giving the Poisson bracket, also the second term of the expansion
involving the
6-th order  derivatives over fermion variables of the product 
$Q_\alpha^{(1)} Q^{\dagger \beta\, (2)}$ contributes in $\{Q_\alpha, Q^{\dagger\beta} \}_{\rm M.B.}$ 
(while $\{Q_\alpha, Q_{\beta} \}_{\rm M.B.}
= \{Q_\alpha, Q_{\beta} \}_{\rm P.B.} = 0$ and the same for $Q^\dagger$ ). 
It is not difficult to see that this extra term is proportional to $\delta_\alpha^\beta$ so that the algebra (\ref{clalg}) 
still holds. The fact that this extra contribution does not vanish means, however, that the Weyl symbol of the quantum
hamiltonian (in contrast to Weyl symbol of quantum supercharges) 
does {\it not} coincide with the classical expression (\ref{HDE}), but has corrections. The situation is the same as in
conventional $\sigma$ models \cite{howto}.  
    
The Weyl ordered quantum supercharges are 
 \be
\label{QDEqu}
 \hat Q_\alpha &=&   \hat P_I f (\gamma_I \xi)_\alpha   + \frac i2  (\partial_J f) ( \gamma_I \xi)_\alpha Q_{IJ} 
-  \frac i{24} \epsilon_{IJKLM} (\partial_I f)  (\sigma_{LM} \xi)_\alpha  Q_{JK} \ , 
  \nn \\
\hat Q^{\dagger\beta} &=&    \hat P_I f (\hat \xi^\dagger  \gamma_I)^\beta    + \frac i2 (\partial_J f)  
(\xi^\dagger  \gamma_I)^\beta   Q_{IJ}
+  \frac i{24} \epsilon_{IJKLM} (\partial_I f)  (\hat \xi^\dagger  \sigma_{LM})^\beta   Q_{JK} \ , 
 \ee
where $\hat P_J$ and $\hat \xi^{\dagger \alpha}$ are differential operators, $\hat P_J = -i \partial_J$, 
 $\hat \xi^{\dagger \alpha} = \partial/\partial \xi_\alpha$.
 The quantum hamiltonian is obtained from the
 anticommutator $\{\hat Q_\alpha, \hat Q^{\dagger\beta} \}$. It can be written
in the form
 \be 
\label{Hqu}
\hat H \ =\ \frac 12 f \hat P_I^2 f - if (\partial_K f) \hat P_J (\hat\xi^\dagger \sigma_{KJ} \xi) \nn \\
- \frac 1{12} \left[ 6(\partial_J f) (\partial_K f) + f(\partial_J \partial_K f) \right] 
(\hat\xi^\dagger \sigma_{JP} \xi) (\hat\xi^\dagger \sigma_{KP} \xi) 
 \ee

We are interested in the case $h = A_5 \equiv \sigma$, $f = 1/\sqrt{\sigma}$.
The wave functions $\Psi$ may depend on $\sigma$, on $A_{1,2,3,4}$, and on holomorphic fermion variables $\xi_\alpha$. 

Let us first find the vacuum states. They should satisfy the conditions
  \be
 \label{QPsi}
 \hat{Q}_\alpha \Psi_{\rm vac} \ = \ 
 \hat Q^{\dagger\beta} \Psi_{\rm vac} \ = \ 0
  \ee
The only solutions to this equation system are 
 \be
\label{123}
\Psi_1 = \sqrt{\sigma}, \ \ \ \ \Psi_2 =  \sqrt{\sigma}\, \xi^4  = \frac {\sqrt{\sigma}}{24} 
\epsilon^{\alpha\beta\gamma\delta} \xi_\alpha \xi_\beta \xi_\gamma \xi_\delta  ,\ \ \ \ 
\Psi_3 =   \sqrt{\sigma}\, {\xi C \xi} \, .
 \ee
All these states are  bosonic. Neither of them  is normalized, however. This could be expected in advance as the allowed 
range $(0, \infty)$ for the variable $\sigma$ implies infinite motion and continuous spectrum 
with not normalizable wave functions.
Note that the normalization integral diverges at $\sigma = \infty$, but not at $\sigma = 0$. In other words,
 the singularity of the moduli space at $\sigma = 0$ does not lead to  trouble  in this case. 

Let us study now  excited  states. Consider first the sector of zero fermionic charge. 
In that case, only the first term
in the quantum hamiltonian (\ref{Hqu}) contributes and the eigenvalue equation is
 \be
-\frac 1{2\sqrt{\sigma}} \left[ \frac {\partial^2}{\partial \sigma^2} + \frac  {\partial^2}{\partial A_M^2}
 \right] \frac 1{\sqrt{\sigma}}
\Psi(\sigma, A_M) \ =\ \lambda \Psi(\sigma, A_M)\ ,
 \ee
where  $M = 1,2,3,4$. The solutions are $\Psi(\sigma, A_M) = g(\sigma) e^{ik_MA_M}$, with $g(\sigma)$ satisfying
 \be
\label{eq4Bes}
 -\frac 1{\sqrt{\sigma}} \left[ \frac {\partial^2}{\partial \sigma^2} - k_M^2 \right] \frac {g(\sigma)}{\sqrt{\sigma}}
 = 2\lambda g(\sigma)\ .
 \ee
Mathematically, this is the textbook problem: the Schr\"odinger equation for the function $g(\sigma)/\sqrt{\sigma}$ 
in  a homogeneous field. The physics
is different, however. First, the physical wave function is still $g(\sigma)$ rather than  $g(\sigma)/\sqrt{\sigma}$. Second, 
the spectral parameter $\lambda$ corresponds not to the energy of the conventional problem, but to the ``field strength'' 
$\partial V/\partial \sigma$. 
Let first $k_M=0$ and assume the energy $\lambda$ to be positive. Then a general  solution to Eq.(\ref{eq4Bes}) is
 \be
\label{PsiF0}
g(\sigma) \ =\  \sigma \left [ A J_{1/3}\left( \frac {2\sqrt{2\lambda}}3 \sigma^{3/2} \right)
+ B J_{-1/3}\left( \frac {2\sqrt{2\lambda}}3 \sigma^{3/2} \right) \right]
 \ee
It seems to be a benign  continuous  spectrum wave function. It vanishes at $\sigma = 0$. For any $A, B$ the normalization integral
$\int d\sigma |g(\sigma)|^2 $ converges at  small $\sigma$ (and diverges  for large $\sigma$, as it should).

However, in a supersymmetric theory, the states should come in multiplets. As we have four different complex supercharges,
 the dimension
of such multiplets should be $2^4 = 16$ in our case. The  states of the multiplet with fermion charge 
$F = 1,2,3,4$ are obtained by the
action of the supercharges on the state with $F=0$. In particular, the four states of unit fermion charge
 associated with the state  (\ref{PsiF0})
are
\be
\label{PsiF1}
\Psi^{F=1}_\alpha \ =\ \hat Q_\alpha  \Psi^{F=0} \ \sim\ 
(\gamma_5 \xi)_\alpha  \sigma \left[ AJ_{-2/3} \left( \frac {2\sqrt{2\lambda}}3 \sigma^{3/2} \right)
- B  J_{2/3} \left( \frac {2\sqrt{2\lambda}}3 \sigma^{3/2} \right) \right]\ .
 \ee
The first term in Eq.(\ref{PsiF1}) behaves as a constant at $\sigma = 0$. We will shortly see that 
non-vanishing of the wave function at the boundary may lead to trouble, but in this particular case it does not. 
The states 
are normalizable and
 admissible,  as the states (\ref{PsiF0}) are.
 Consider now the states $\hat Q_\alpha \hat Q_\beta \Psi^{F=0} $ in the sector $F=2$. 
The state $\propto A$ in Eq. (\ref{PsiF0})
leads to the following six states,
 \be
\label{PsiF2}
\Psi^{F=2} \ \sim \ (\xi  C \xi)\, \sigma J_{1/3}\left( \frac {2\sqrt{2\lambda}}3 \sigma^{3/2} \right);\ \ \ 
\sim \    (\xi  C \gamma_M \xi) \, \sigma J_{1/3}\left( \frac {2\sqrt{2\lambda}}3 \sigma^{3/2} \right);\nn \\
(\xi C \gamma_5 \xi) \, \sigma J_{-5/3}\left( \frac {2\sqrt{2\lambda}}3 \sigma^{3/2} \right) \ .
 \ee
and the state    $\propto B$ leads to the states of the same structure, but with the sign of the
 Bessel indices reversed. 
We observe that five of six states in (\ref{PsiF2}) are benign, but the 
sixth $\propto$ $\xi C  \gamma^5 \xi$ behaves as
$\sigma^{-3/2}$ at small $\sigma$, is not renormalizable there, and  not admissible by that reason.
 But if we want to keep supersymmetry, we
cannot reject only    one  state in the 
supersymmetric 16-plet. We should also throw fifteen others away ! 
 \footnote{The observation that, when defining Hilbert space in SUSY theories, one has to filter 
out not only the ``bad'' 
(nonrenormalizable, not gauge-invariant etc.) states, but also the superpartners of such states even though these 
superpartners
look benign  by themselves was made in \cite{SSV}. We refer the reader to that paper for more examples 
and discussion.}    

On the other hand, all the states stemming from the state $\sim \sigma J_{-1/3}(z)$ in (\ref{PsiF0}) ---  the states
 \be
\label{16states}
  F=0: \ \ \ \ \ &\sim& \sigma J_{-1/3}(z);  \nn \\
F=1: \ \ \ \ \ & \sim& ( \gamma_5  \xi )_\alpha\, \sigma J_{2/3}(z); \hfill  \nn \\
F=2: \ \ \ \ \ &\sim& (\xi  C \xi)\, \sigma J_{-1/3}(z),\ \ \sim (\xi  C  \gamma_M \xi)\, \sigma J_{-1/3}(z),\ \
 \sim (\xi C  \gamma_5 \xi)\, \sigma J_{5/3}(z); \hfill  \nn \\
 F=3: \ \ \ \ \ &\sim& ( \ \gamma_5)_\beta^\alpha \epsilon^{\beta\gamma\delta\epsilon} 
\xi_\gamma \xi_\delta \xi_\epsilon \,\sigma J_{2/3}(z); 
\hfill   \nn \\
F=4: \ \ \ \ \  &\sim&  \xi^4 \, \sigma J_{-1/3}(z)\ . \hfill
 \ee
($z = 2\sqrt{2\lambda}\,  \sigma^{3/2}/3$), obtained from each other by the action of supercharges (\ref{QDEqu}) 
are admissible and are present in the supersymmetric spectrum of our system.

     What happens if $\lambda$ is negative ? If limiting ourselves with the sector $F=0$, one obtains 
the solutions
 \be
\label{MacF0}
g(\sigma)\  \sim \ \sigma K_{1/3} \left( \frac{ 2 \sqrt{-2\lambda}}3  \sigma^{3/2} \right)\ .
 \ee
 Being normalizable not only at $\sigma = 0$, but also at infinity, they seem to be quite respectable.
But the presense of the states with negative energy is in obvious contradiction with supersymmetry. 
Indeed, there are at least 
 two  reasons  to reject them. 
 \begin{enumerate}
\item
First, recall the standard  proof that the energies of all the  states in a supersymmetric
system are positive or zero.
We have
 \be
\langle \Psi|\hat H| \Psi \rangle \ =\ \langle \hat Q \Psi| \hat Q\Psi \rangle +  
\langle \hat Q^\dagger \Psi| \hat Q^\dagger \Psi \rangle \geq 0\, .
 \ee
But this is based on the assumption that $\hat Q^\dagger$ is adjoint to $\hat Q$. One can note, however, that this property does not hold
if including in the spectrum the state (\ref{MacF0}). Indeed, only the first term in the expression (\ref{QDEqu}) for $\hat Q_\alpha$
acts nontrivially on the states in the sector $F= 0$. Disregarding the irrelevant spinor structure, consider the operators 
\be
\hat S = \frac 1{\sqrt{2}} \partial_J f,\ \ \ \ \ \hat S^\dagger =
 -\frac 1{\sqrt{2}} f\partial_J;\ \ \ \ \ \ \hat H' =  \hat S^\dagger \hat S\ .
 \ee
Then the identity  
 \be
\label{HSS}
\langle g|\hat H'|g \rangle = -\frac 12 \int_0^\infty d\sigma\, g(\sigma) \frac 1{\sqrt{\sigma}}
 \frac {d^2}{d\sigma^2}  
\frac {g(\sigma)}{\sqrt{\sigma}}  = \int_0^\infty d\sigma\, [\hat S g(\sigma))]^2 > 0 
 \ee
would be correct if the boundary term 
 \be
 \label{bterm}
\left.  \frac 12 \frac {g(\sigma)} {\sqrt{\sigma}} \frac d {d\sigma} \, \left( \frac {g(\sigma)} {\sqrt{\sigma}} \right) 
\right|_{\sigma = 0} 
  \ee
vanished. But it does not ! One can be convinced that the contribution (\ref{bterm}) to 
$\langle g|\hat H'|g \rangle$ is negative, which overcomes the positive contribution of the R.H.S. of
Eq.(\ref{HSS}) and allows for the eigenvalue of $\hat H'$ to be negative.  

Thus, if we want to keep $\hat Q_\alpha$ and $\hat Q^{\dagger\beta}$ 
conjugate to each other and $\hat H$ hermitian, the state (\ref{MacF0}) should be rejected.

\item The states in the sector $F=2$ derived from 
the state (\ref{MacF0})
by the action of supercharges include the state $\sim (\xi C \gamma_5 \xi)\, 
\sigma K_{5/3} ( 2 \sqrt{-2\lambda}\, \sigma^{3/2}/3)$, which  is not normalizable at the origin and not admissible. 
Hence, the whole  lame would-be multiplet should be rejected, as discussed above.   
 \end{enumerate}

Up to now we have only discussed the states with $k_M = 0$. Let us consider now  
 the case of nonzero $k_M$. Let for definiteness $k_1 \equiv k \neq 0, k_{2,3,4} = 0$. The solutions with negative $\lambda$
should surely be rejected by the argument 1 above. If $\lambda$ is positive, the solutions to the equation (\ref{eq4Bes}) change their nature
at the point $\sigma_0 = k^2/2\lambda$, they represent oscillatory Bessel functions for $\sigma \geq \sigma_0$ and modified Bessel functions 
(related to Airy functions)
at $0 \leq \sigma \leq \sigma_0$. To find whether a state is admissible or not, we have to study the behaviour of the
solution at the vicinity of $\sigma = 0$ where a generic solution to Eq. (\ref{eq4Bes}) has the form
 \be
\label{I13}
\Psi^{F=0} \equiv g(\sigma) \  =\ \sqrt{\sigma} \left[ A\sqrt{\sigma_0 - \sigma}\,  I_{1/3}  
\left( \frac {2 \sqrt{2\lambda}}3  (\sigma_0 - \sigma)^{3/2} \right) \right. \nn \\ 
\left.  +  B\sqrt{\sigma_0 - \sigma} I_{-1/3}  \left( \frac {2 \sqrt{2\lambda}}3 \,  (\sigma_0 - \sigma)^{3/2} \right)\right]\ .
 \ee
 The superpartners of this state in the sector $F=1$ are 
 \be
 \label{kF1}
 \Psi_\alpha^{F=1} \ =\ \hat Q_\alpha \Psi^{F=0} \ =\ \frac {kg(\sigma)}{\sqrt{\sigma}} (\gamma_1 \xi)_\alpha -
 i \frac d{d\sigma} \left( \frac {g(\sigma)}{\sqrt{\sigma}} \right)
(\gamma_5 \xi)_\alpha 
 \ee
and
  \be
 \label{kF2}
 \Psi_{\alpha\beta}^{F=2}  = \hat Q_\alpha \hat Q_\beta \Psi^{F=0} = \frac {k^2g(\sigma)}{\sigma} (\gamma_1 \xi)_\alpha(\gamma_1 \xi)_\beta
  -  \frac d{d\sigma} \left( \frac 1{\sqrt{\sigma}}  \frac d{d\sigma} \frac {g(\sigma)}{\sqrt{\sigma}} \right)  
(\gamma_5 \xi)_\alpha(\gamma_5 \xi)_\beta  \nn \\ 
- \frac 1{4\sigma^{3/2}}  \frac d{d\sigma} \left ( \frac {g(\sigma)}{\sqrt{\sigma}} \right) (\gamma_M \xi)_\alpha(\gamma_M \xi)_\beta
+ \frac 1{24\sigma^{3/2}}  \frac d{d\sigma} \left ( \frac {g(\sigma)}{\sqrt{\sigma}} \right) 
(\sigma_{MN} \xi)_\alpha (\sigma_{MN} \xi)_\beta\ .
 \ee
(the summation runs over $M,N = 1,2,3,4$).
For generic $A,B$, the projection of the wave function (\ref{kF2}) on the structure $\sim \xi C \gamma_5 \xi $ behaves as
$\sigma^{-3/2}$ at small $\sigma$ and is not normalizable. But for one particular choice of the ratio $A/B$, 
this leading singularity vanishes and
the wave function behaves at the origin as $\sim \sigma^{-1/2}$, which is ``almost normalizable''. 
 The members of this supermultiplet with fermion charge $F=3$ and $F=4$ can be obtained by duality transformation 
from  the states in the sectors
$F=1$ and $F=0$, respectively and have the same $\sigma$ dependence. 
This is best seen by noting that the states (\ref{kF2}) can be alternatively obtained by acting with 
$\hat Q^{\dagger \alpha}\hat Q^{\dagger \beta} $  on the state
$g(\sigma) \xi^4 $.

To include in the supersymmetric spectrum 
 the multiplets involving
the states, where the normalization integral  diverges logarithmically at the origin, or to reject them 
 is a matter of taste and convention. We believe that it is more natural to include them. In this case,  
 the full spectrum involves the 16-plets labelled by five quantum numbers $(\lambda \geq 0, k_{M})$. 
This corresponds to the presence of five
bosonic dynamic variables in our  system.  

\section{Discussion}
We can return now to the question posed at the end of Sect. 2. As far as the dimensionally reduced system (\ref{Lpredel})
is concerned, the   answer is definitely positive --- in spite of the presense of singularity at $\sigma = 0$, the 
{\it supersymmetric} spectrum of the theory is nontrivial and bonded by zero from below, as it should. In other words, the situation
is similar, indeed, to the QM problem with potential $\sim -\gamma/r^2$ for small $\gamma$. Though classical 
trajectories are singular, the quantum problem is well defined. We would {\it not} be able, however, to define the quantum problem
in this case without invoking supersymmetry. Ghost states having unbounded negative energies  do exist as solutions of the
Schr\"odinger equation, and only the fact that they do not have normalizable superpartners, allows one to reject them.

 Maybe this is the most important observation of the paper. We suggest that the same is true in the TOE (representing,
 according to our hypothesis \cite{TOE}, a field theory in higher dimensions with higher derivatives). 
One can speculate that, though ghost states
associated with higher derivatives are formally present in the spectrum, the {\it physical} 
Hilbert space ${\cal H}$ of such a theory involves only supermultiplets of the states with positive energies. Now, ${\cal H}$
 should be closed such that when acting by physically admissible operators on any state $\Psi \in {\cal H}$, we stay within
 ${\cal H}$ (in more physical language, ghosts are not created in collisions of usual particles). We tend to 
prefer the mechanism of getting
rid of the ghosts discovered in this paper to the  mechanism unravelled in \cite{duhi} --- a QM higher-derivative model considered
there was not supersymmetric and the bottom in the spectrum appeared by the reasons not related to supersymmetry. The
arguments based on supersymmetry have more aesthetic appeal and have the advantage of being {\it universal}. On the other hand, 
the ghost-free $QM$ system considered in \cite{duhi} has one important common feature with the system (\ref{Lpredel}). 
In both theories, the spectrum is essentially nonperturbative.  
   Obviously, further studies of this question are necessary.

We can conjecture that the hamiltonian has a well defined nonperturbative spectrum not only for  the reduced system, but also for
the field theory (\ref{L5D}) in the sector with zero scalar expectation value $\langle \sigma \rangle = 0$.  As this spectrum
is essentially not perturbative and we do not have analytic tools to study field theory nonperturbatively, we cannot say much about
the nature of this spectrum and about the quantum dynamics of this theory.  The only thing that we still can suggest 
(based on the absence of the scale parameter in the lagrangian) is that
this dynamics {\it is} nontrivial and stays nontrivial in continuum limit. On can recall in this respect $2D$ conformal theories. 
Like the theory (\ref{L5D}), they do not have a scale parameter. They do not have well defined asymptotic states and the scattering
matrix cannot be defined (on the other hand, one can define $S$-matrix  for  {\it deformed} conformal theories, like the
Sine-Gordon model or Ising model at critical temperature and nonzero magnetic field \cite{Zam}). $2D$ conformal
  theories are known to have nontrivial interesting dynamics. We do not see reasons why the $5D$ 
 conformal theory considered in this paper should not have one.

I am indebted to E. Ivanov for fruitful discussions.

 \section*{Appendix}
For references purposes, we will give here some useful formulae referring to the component formulation of the 
generic DE model. The Euclidean $\gamma$ matrices satisfy 
$$\gamma_I^\dagger = \gamma_I,\ \ \ \ \ \gamma_I \gamma_J + \gamma_J \gamma_I = 2\delta_{IJ}\,,\ \ \ \ \ 
 C\gamma_I^T = \gamma_I C\, .$$ 
The matrices $C$ and $\Gamma_I = C\gamma_I$ are antisymmetric while $\Gamma_{IJ} = C\sigma_{IJ}$ are 
symmetric in the 
spinor indices. One of the possible
explicit representations is
 $$ \gamma_{1,2,3} = \sigma_{1,2,3} \otimes \sigma_3\, , \ \ \ \gamma_4 = 1\!\!1 \otimes \sigma_1\, , \ \ \  
\gamma_5 = 1\!\!1 \otimes \sigma_1\, , \ \ \
C = i\sigma_2 \otimes \sigma_1\, .$$ 
 Let us  write the lagrangian in the form analogous to (\ref{L5D}) so that the $SU(2)$ R-symmetry of the model
 would be explicitly
seen. We have
 \be
\label{DESU2}
 L \ =\ \frac h2 \left( \dot A_I^2 - i \psi^k C \dot \psi_k \right) + \frac i4 (\partial_J h) \dot A_I \psi^k 
\Gamma_{JI} \psi_k 
- \frac h4 D^{jk} D_{jk} + \nn \\ 
\frac 14 D^{jk} (\partial_I h) \psi_j \Gamma_I \psi_k - \frac 1{24} (\partial_I \partial_J h) 
(\psi^j \Gamma_I \psi^k) 
 (\psi_j \Gamma_J \psi_k)\ .
  \ee
This lagrangian is unvariant up to a total derivative with respect to the SUSY transformations
 \be
\delta A_I &=& E^k \Gamma_I \psi_k \, ,\nn \\
\delta \psi_j &=& - i \dot A_I \gamma_I E_j + D_{jk} E^k\, ,  \nn \\
 \delta D_{jk} &=& i \dot \psi_j C E_k + i \dot \psi_k C E_j\, . 
 \ee
Expressing out the auxiliary fields and using the substitutions (\ref{sootv}), we reproduce the result
(\ref{DEcomp}) above.

Finding {\it the} total derivative in $\delta L$ represents a technically most difficult part in deriving 
the expression for the supercharges $Q^j$.
It can still be done by using various Fierc identities, for example
 \be
(\psi^j \Gamma_{(I} \psi^k)\,  (\psi_j \Gamma_{J)} \psi_k) \ =\ \frac 12 (\psi^j \Gamma_{IK} \psi_j)\,  
\psi^k \Gamma_{JK} \psi_k)  - \frac 18
\delta_{IJ} (\psi^j \Gamma_{KL} \psi_j )^2\ ,
 \ee
the harmonicity of $h$, and the useful identity
 \be 
\label{eps5}
\epsilon_{IJKLN} S_{MN} + \epsilon_{JKLMN} S_{IN} + \epsilon_{KLMIN} S_{JN} + \epsilon_{LMIJN} S_{KN} + 
\epsilon_{MIJKN} S_{LN} \nn \\
=  \epsilon_{IJKLM} S_{NN} 
 \ee
valid for any tensor $S_{IJ}$. We obtain finally (note that the contributions involving auxiliary fields cancel)
 \be
\label{Qj}
 Q^j_\alpha \ =\ P_I (\gamma_I \psi^j)_\alpha - \frac {i\partial_J h}8 (\psi^k \Gamma_{IJ} \psi_k )\,
 (\gamma_I \psi^j)_\alpha
+  \epsilon_{IJKLM} \frac {i \partial_I h}{96}   (\psi^k \Gamma_{JK} \psi_k)\,   (\sigma_{LM} \psi^j)_\alpha\ ,
 \ee
which gives (\ref{QDE}).

\end{document}